\documentstyle[prb,aps,epsfig,multicol,epsf,amssymb]{revtex}
\begin{document}
\draft
\title{Coherence in magnetic quantum tunneling}
\author{Julio F. Fern\'andez}
\address{Instituto de Ciencia de Materiales de Arag\'on\\Consejo Superior de 
Investigaciones Cient\'{\i}ficas\\and Universidad de Zaragoza, 50009-Zaragoza, Spain\\} 
\date{\today}
\maketitle
\vspace{0.3cm} 
\begin{abstract}                
Currently, spin tunneling at very low temperatures is assumed
to proceed as an incoherent sequence of events that take place
whenever a bias field $h(t)$ that varies randomly with time $t$ becomes
sufficiently small, as in Landau-Zener transitions. 
We study the behavior of a suitably defined coherence time $\tau_\phi $.
Coherence effects become significant when $\tau_\phi \gtrsim \tau_h$,
where $\tau_h$ is the correlation time for $h(t)$. 
The theory of tunneling of Prokof'ev and Stamp (PS), which rests on the
assumption that $\tau_\phi\lesssim \tau_h$, is extended beyond this
constraint. It is shown, both analytically and numerically, that
$\tau_\phi\gtrsim\tau_h$ when $\tau_h\delta h\lesssim \hbar$,
where $\delta h$ is the rms deviation of $h$. Equations
that give $\tau_\phi$ and the tunneling rate as a function
of $\tau_h\delta h$ both for $\tau_h\delta h\gtrsim\hbar$,
where the theory of PS hold, and for $\tau_h\delta h\lesssim\hbar$,
where it does not, are derived.

\end{abstract}  
\begin{multicols}{2}
\narrowtext
Magnetic quantum tunneling (MQT) of large spins has become a subject of much interest.
Much of it stems from the expectation
that MQT will contribute to our understanding of decoherence.\cite{nature1}
Decoherence effects play an important role in the transition from quantum
to classical physics \cite{zurek}
and in quantum computing.\cite{unruh}
Magnetic quantum tunneling (MQT) through thermally activated states
is thought to proceed incoherently.\cite{luis2,julio} However, 
at sufficiently low temperatures, tunneling must proceed
though the ground state doublet\cite{wern0}. Two-state systems are often used to
model tunneling under such conditions.\cite{garg,stamp1}
Existing theory treats tunneling between these two states as an {\it incoherent}
sequence of attempts that take place whenever the difference between the
two-state energies becomes sufficiently small.\cite{stamp2}
It is in the spirit of Landau-Zener
schemes that are often used in the theory of non-adiabatic transitions
to estimate transfer rates between two states whose energies sometimes cross.\cite{nikitin}

Here, I study coherence effects in MQT by means of
the following two-state model,
\begin{equation}
{\cal H}=\varepsilon\sigma_z+\Delta\sigma_x-h(t)\sigma_z,
\label{hamilton}
\end{equation}
where ${\cal H}$ is the Hamiltonian, $\sigma_z$ and $\sigma_x$ are
Pauli spin matrices, $\varepsilon$ is an energy that is constant in time,
$h(t)$ varies with time $t$ following
a stationary random Gaussian process,
and $\Delta$ stands for the tunneling frequency
($\hbar =1$, and $k=1$, where $k$ is Boltzmann's constant, from here on).
Two-state systems have been used to study MQT,\cite{garg,stamp1,stamp2} 
but they have also been used to study decoherence in other systems,
\cite{leggett} such as in (hypothetical) quantum computers.\cite{unruh}
The heat bath is replaced by $h(t)$ in this model. Consequently, the model
is applicable only if $\Delta \ll T$ and $\varepsilon\ll T$, where $T$ is the temperature.
This is amply satisfied in present day experiments in MQT
when no transverse field is applied. 

The aim of this paper is to study within the model defined above
the behavior of the coherence time $\tau_\phi $
(defined below) and of the tunneling rate $\Gamma$ as a function
of $\tau_h\delta h$, where
$\delta h$ is the rms deviation of $h(t)$ and $\tau_h$ is its correlation time.
Unless stated otherwise,
$\tau_h\Delta\ll 1$, $\Delta\ll\delta h$,
and $\Delta\ll(\tau_h\delta h)\delta h$
are assumed to hold everywhere. It is shown that decoherence effects are partially
averaged out when $\tau_h\delta h\lesssim 1 $. 
I derive, and support with
numerical results, that if $\tau_h\delta h\lesssim 1 $, then
$\tau_\phi /\tau_h\approx 1/(\tau_h\delta h)^2$,
instead of $\tau_\phi \approx 1/\delta h$ that is shown to hold if $\tau_h\delta h\gtrsim 1 $.
Furthermore, instead of the relation,
\begin{equation}
\Gamma \simeq \sqrt{\pi \over 2}{{\Delta^2}\over{\delta h}}\exp\left[-
{{\varepsilon^2}\over{2(\delta h)^2}}\right],
\label{incoh}
\end{equation}
derived below for the tunneling rate under conditions of full incoherence (that is,
for $\tau_\phi\lesssim\tau_h$), which is
the underlying assumption in the spin tunneling theory of Prokof'ev and Stamp,
\cite{stamp1} the relation 
\begin{equation}
\Gamma \simeq {{\Delta^2\tilde\tau_{\phi}}\over{1+\varepsilon^2\tilde\tau_{\phi}^2}},
\label{coh}
\end{equation}
where $\tilde\tau_{\phi}^{-1}\equiv 2\tau_h(\delta h)^2$, is derived
under the condition that $\tau_h\delta h\lesssim 1 $.
Note that $\tilde\tau_{\phi}\approx\tau_\phi$ if $\tau_h\delta h\lesssim 1 $.
Finally, concluding remarks about the likelyhood that $\tau_h\delta h\lesssim 1$
is realized in Fe$_8$ and Mn$_{12}$ are made.

In order to define a measure of coherence,\cite{referee1}
consider the probability $p_\sigma(t)$
that the system evolve in time $t$ from an intial
state $|\psi \rangle$ into a final state $|\sigma\rangle$. We can then write,
\begin{equation}
p_\sigma(t+\tau)=\sum_{\sigma^\prime ,\sigma^{\prime\prime}}
w(\sigma ,\sigma^\prime ,\sigma^{\prime\prime};\tau )
\langle \sigma^\prime |\psi (t)\rangle\langle\psi(t) |\sigma^{\prime\prime}
\rangle ,
\label{smeq}
\end{equation}
where $w(\sigma ,\sigma^\prime ,\sigma^{\prime\prime};\tau )=
\langle \sigma^{\prime\prime}|U^\dagger({\{ h\}},\tau)|\sigma\rangle
\langle \sigma |U({\{ h\}},\tau)|\sigma^\prime \rangle$,
$U({\{ h\}},\tau)={\cal T}\exp [-i\int_0^\tau {\cal H}(t_1)dt_1]$ for a given $h(t)$ in
Eq. (\ref{hamilton}), and ${\cal T}$ is a time ordering operator.
Averaging Eq. (\ref{smeq}) over different $h(t)$ time sequences, then
when $\tau_h$ for $h(t)$ fulfills $\tau_h\ll \tau $, $t$,
its right hand side breaks up into sums of {\it products of two terms},
$\langle w(\sigma ,\sigma^\prime ,\sigma^{\prime\prime};\tau )\rangle_h$ and
$\langle\langle \sigma^\prime |\psi (t)\rangle\langle\psi(t) |\sigma^{\prime\prime}
\rangle\rangle_h $,
where $\langle\rangle_h$ stands for average over all different histories
of $h(t)$. We are interested here in
establishing the physical conditions under which
$\langle w(\sigma ,\sigma^\prime ,\sigma^{\prime\prime};\tau )\rangle_h\rightarrow
\delta_{\sigma^\prime,\sigma^{\prime\prime}}W_{\sigma,\sigma^\prime}(\tau )$
ensues, because this leads to
\begin{equation}
p_\sigma(t+\tau)=\sum_{\sigma^\prime}W_{\sigma,\sigma^\prime}(\tau )
p_{\sigma^\prime}(t).
\label{pauli}
\end{equation}
No interference terms from Eq. (\ref{smeq}) appear in this equation. Thus,
probabilities, rather than probability amplitudes, for allternative paths to a final
state are summed. Its applicability is, therefore, tantamount to incoherence.
Solutions to Eq. (\ref{pauli}) exhibit no oscillations.
For any two state system, they relax to the final state exponentially.
Consider the off-diagonal quantity,
\begin{equation}
q(\sigma,\sigma^\prime ,\sigma^{\prime\prime} ;\tau)
={{\langle w(\sigma ,\sigma^\prime ,\sigma^{\prime\prime};\tau)\rangle_h}
\over {(p_1p_2)^{1/2}}}
\label{qq}
\end{equation}
for $\sigma^\prime\neq \sigma^{\prime\prime}$,
where $p_1=\langle |\langle\sigma^{\prime\prime} |U^\dagger({\{ h\}},\tau)
|\sigma\rangle |^2\rangle_h$, and
$p_2=\langle |\langle\sigma |U({\{ h\}},\tau)
|\sigma^{\prime}\rangle |^2\rangle_h$. As shown above, $q=0$ leads to Eq. (\ref{pauli}),
and therefore to incoherence. 
On the other hand, the system evolves in time in a pure state if there is no averaging
over different $h(t)$ time evolutions. No decoherence occurs then, and
it follows from  the definition of $q$ that $\mid q\mid =1$ then. Quantity
$q$ provides, therefore, an appropriate measure of coherence between the two alternative
paths $\sigma^\prime$ and $\sigma^{\prime\prime}$. This is in analogy to an
interference experiment that may be performed to test the degree of coherence between the
parts of a particle-wave that go through two slits placed at $\sigma^\prime$ and
$\sigma^{\prime\prime}$. Let the {\it coherence time} be defined as the
smallest time $\tau_\phi $ such that, say, $\mid q(\tau)\mid <1/2$ for all $\tau >\tau_\phi $.
(Below, $q$ is independent of
$\sigma,\sigma^\prime ,\sigma^{\prime\prime}$ throughout. Accordingly,
these variables of $q$ are omitted from here on.)

Numerical results are obtained for the system's time evolution
for a given $h(t)$ time sequence by iterating a finite difference
Schr\"odinger's equation,
\begin{equation}
\psi (t+Dt)=\{1-i Dt{\cal H}(t)-[Dt{\cal H}(t)]^2/2\}\psi (t),
\end{equation}
where ${\cal H}$ is defind in Eq. (\ref{hamilton}), $\psi$ is a two-component
state function, and $Dt$ is a suitably small time interval
($Dt\lesssim 10^{-6}/\Delta$ keeps the normalization condition
satisfied within 1 part out of 1000 throughout the reported simulations). 

The values of $h(t)$ are assumed to follow from a stationary random Gaussian
process, with a distribution for $h$ centered on $0$, and
$h(t)$ is correlated over time according to 
$\langle h(t)h(0)\rangle_t=(\delta h)^2\exp (-t/\tau_h)$.
This is accomplished with Langevin's (finite difference) equation,
$h(t+Dt)=h(t)-Dt[h(t)/\tau_h +f(t)]$, where $f(t)$ is assigned an independent
random number at each value of $t$. All values of $\tau_h$ and $Dt$ used
fulfill $Dt \leq \tau_h/10$. The values of $h$ obtained follow
the prescribed behavior for it if $\langle f\rangle =0$ and
$\langle f(t)f(t^\prime)\rangle =\delta_{t,t^\prime}(2-Dt/\tau_h)
(\delta h)^2/(\tau_hDt)$,
where $\delta_{t,t^\prime}$ is the Kronecker delta function.
The correction term $Dt/\tau_h$ comes about from applying the 
fluctuation dissipation theorem to the {\it finite difference}
Langevin equation used here.

Numerical results obtained for the tunneling probability and for $q$
are shown in Fig. 1 for $\delta h/\Delta =50$ and $\tau_h\Delta=10^{-4}$.
Note that even though $\delta h$ is 50 times larger
than $\Delta$, the tunneling probability undergoes oscillations,
and, correspondingly, $q\ll 1$ only after a time
that is much larger than $\Delta^{-1}$.
As shown below, the high degree of coherence exhibited in Fig. 1
occurs when $\tau_h(\delta h)^2 \lesssim \Delta$. It may be realized in
MQT experiments by enlarging $\Delta$, through the application of an external field,
as in Refs.,\cite{tejada,gatto} where $\Delta\approx 30$K is several
orders of magnitude larger than the linewidth [see Eq. (\ref{coh}) below].

\begin{center}
\begin{figure}
\epsfig{file=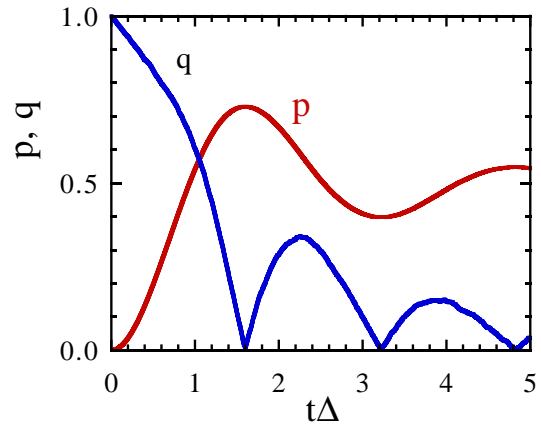,width=8cm}
\caption{The probability $p$ that tunneling occurs in time $t$ and the
off-diagonal quantity $q$ are shown versus $t\Delta$.
The data points shown follow from an average over $4\times 10^3$ time
evolutions with different realizations of $h(t)$,
with a vanishing time average value of $h$,
$\delta h/\Delta =50$, and $\tau_h\Delta=10^{-4}$.} 
\end{figure}
\end{center}

This extreme condition is worth considering, because it serves to illustrate
how decoherence effects can be drastically supressed. Quite simply,
rapid fluctuations in $h$ are almost completely averaged out. More specifically,
they are scaled down by the factor $\sqrt{\tau_\phi /\tau_h}$ in time $\tau_\phi $.
Thus, $\delta h/\Delta =50$ is scaled down by $100$ in Fig. 1.

Numerical results that illustrate the effects of partially averaging decoherence out
are exhibited in Fig. 2.
\begin{center}
\begin{figure}
\epsfig{file=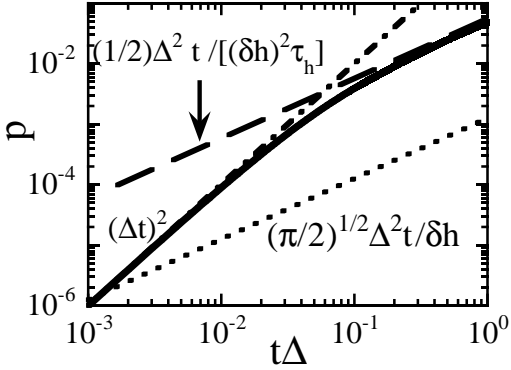,width=8cm}
\caption{Tunneling probability $p$ versus time. The curves shown follow
from an average over $4\times 10^3$ time
evolutions with different realizations of $h(t)$, $\varepsilon =0$,
$\delta h/\Delta =10^3$, and $\tau_h\Delta=10^{-5}$.
From data points obtained for $q$ (not shown), the value
of time where $q=1/2$, that is $\tau_\phi $, can be obtained. It turns
out that $\tau_\phi\simeq 0.07/\Delta$.
The full line is for $p$; the dotted line
is for the prediction that follows for $p$ assuming incoherence for times
shorter than $\tau_h$; the dashed-dotted line is for $(\Delta t)^2$,
that is fully coherent evolution of $p$; finally, the long dashed line is
for the assymtotic evolution after decoherence sets in.} 
\end{figure}
\end{center}
To start the derivation of the tunneling rates,
consider the solution to Schr\"odinger's equation,
$\psi(t)=U_0(\{h\},t)U_I(\{h\},t)\psi(0)$,
where $U_0(\{h\},t)=\exp {[-i\int_0^t{\cal H}_0(\tau)d\tau ]}$,
${\cal H}_0(\tau )=[\varepsilon -h(\tau )]\sigma_z$,
$U_I(\{h\},t)={\cal T}\exp {[-i\int_0^tV_I(\tau)d\tau ]}$,
$V_I(t)= U_0^\dag(\{h\},t)VU_0(\{h\},t)$, and $V=\Delta\sigma_x$.
The probability amplitude $a(t)$ that tunneling take place in time $t$, that is, that a
transition from state $\sigma_z=-1$ to state $\sigma_z=+1$ occurs, is
\begin{equation} 
\mid a(t)\mid =\mid \int_0^t dt_1\;\langle 1|\Delta\sigma_x e^{-2i\int^{t_1}_0{\cal H}_0(\tau)d\tau}|-1\rangle\mid ,
\label{aoft}
\end{equation}
in first order perturbation theory, where use has been made of the relation
$\exp (A\sigma_z)\;\sigma_x=\sigma_x\exp (-A\sigma_z)$ for any $c$-number $A$. Now, since
${\cal H}_0(\tau )\mid\pm 1\rangle =\pm [\varepsilon -h(\tau )]\mid\pm 1\rangle$,
it follows that $\int^{t_1}_0{\cal H}_0(\tau)d\tau$ can be replaced in Eq. (\ref{aoft}) by
$[\varepsilon-H(t_1)]t_1$, where $H(t)$ is defined by 
$t H(t)=\int_0^t d\tau^\prime h(\tau^\prime )$.

In order to get some feeling for the physics behind the results to be derived,
a rough argument that leads to semiquantitative relations
for the coherence time and the tunneling rate $\Gamma$ is given first.
For shortness, the argument is restricted to the condition $\varepsilon =0$,
but the derivation itself of Eqs. (\ref{incoh}) and (\ref{coh}) is not.
It follows from Eq. (\ref{aoft}), and the definitions of $q$ and of $\tau_\phi$ that,
in general,
\begin{equation}
\tau_\phi\approx 1/\delta H(\tau_\phi ),
\label{dH}
\end{equation}
where $\delta H(\tau_\phi )$ is the rms value of $H(\tau_\phi )$. 
Furthermore, $\delta H(\tau_\phi )\approx \delta h$ if
$\tau_h\delta h\gtrsim 1$. Therefore,
\begin{equation}
\tau_\phi\approx 1/\delta h,
\label{tauc1}
\end{equation}
if $\tau_h\delta h\gtrsim 1$. There is then complete incoherence, that is,
$\tau_\phi\lesssim \tau_h$. The scheme of Prokof'ev and Stamp\cite{stamp1},
of an {\it incoherent} sum of tunneling probability increments
that occur whenever $h(t)$ becomes comparable to $\Delta$,
is then justified. If, on the other hand, $\tau_h\delta h\lesssim 1$, then
$\delta H(\tau_\phi )\approx \delta h/\sqrt{\tau_\phi/\tau_h}$, since
$\tau_\phi/\tau_h$ is, in a rough sense, the number of times
that $h$ fluctuates independently during time $\tau_\phi$. 
Therefore, substitution into
Eq. (\ref {dH}) gives
\begin{equation}
\tau_\phi /\tau_h \approx 1/(\tau_h\delta h)^2
\label{tauc2}
\end{equation}
if $\tau_h\delta h\lesssim 1$. This equation exhibits explicitely
how averaging out of
fluctuations makes $\tau_\phi$ larger than $\tau_h$ if
$\tau_h\delta h\lesssim 1$. Data points obtained numerically for
$\tau_\phi$ are shown in Fig. 3. Note that Eq. (\ref{tauc2}) also implies
that coherence times become larger
than tunneling times when $\tau_h\delta h^2\lesssim\Delta$.
\begin{center}
\begin{figure}
\epsfig{file=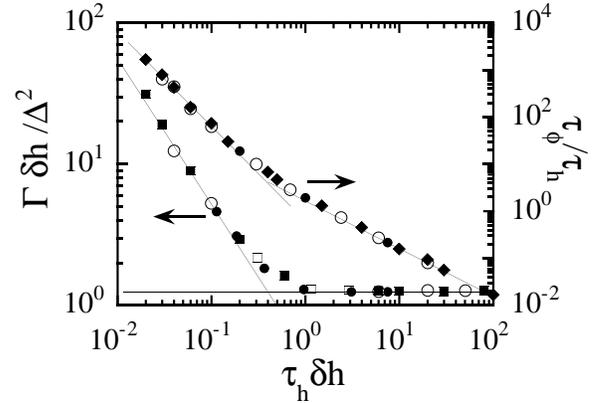,width=8cm}
\caption{Data points from numerical time
evolutions with $\varepsilon =0$ for $\tau_\phi /\tau_h$ and for
for $\Gamma\delta h/\Delta$
versus $\tau_h\delta h$. $\blacksquare$, and $\blacklozenge$ stand for
$\delta h=1000$, $\circ$, $\bullet$, and $\Box$
stand for $300, 75$, and $30$, respectively.
Each data point for $\tau_\phi /\tau_h$
and for $\Gamma\delta h/\Delta$, stands for an average
over at least $10^3$ and $10^4$ $h(t)$ time evolutions, respectively. 
The straight lines for $\tau_\phi /\tau_h$ are guides to the eye, with slopes
$1$ and $2$; the straight lines for $\Gamma\delta h/\Delta$
follow from Eqs.
(\ref{incoh}) and (\ref{coh}). All data points shown are for values of $\tau_h$
that fulfill $\tau_h\Delta\lesssim 0.1$ and $\tau_h(\delta h)^2/\Delta\gg 1$.
Error bars are covered by the symbols for the data points.}
\end{figure}
\end{center}
The tunneling rate $\Gamma$ is estimated next for $\varepsilon =0$
and $\tau_h\delta h\lesssim 1$. It follows from Eq. (\ref{aoft}) that
the probability $p[t,H(t)]$ for a given $H(t)]$ that
a spin has tunnelled at time $t$ is 
$p[t,H(t)]\approx \Delta^2t^2$ for $tH(t)\lesssim 1$.
It may seem strange that a spin under a large field tunnels as fast
as one for which $H=0$ even though the line shape is known to be a Lorentzian
function, but note that
$\Delta^2(\Delta^2+H^2)^{-1}\sin^2(\Omega t)$, where   
$\Omega^2=H^2+\Delta^2$, is approximately $\Delta^2t^2$ for $\Omega t\lesssim 1$,
wich is independent of $H$. Averaging over all $h(t)$ sequences leads to 
$p(\tau_\phi)\approx\Delta^2\tau_\phi^2$. Equation (\ref{pauli}) gives
$p(t)\approx p(\tau_\phi)t/\tau_\phi$ for $t\gtrsim\tau_\phi$, since
$\tau_h\lesssim \tau_\phi$ under the assumed condition
$\tau_h\delta h\lesssim 1$. It follows then that, 
$\Gamma\sim\Delta^2/[\tau_h(\delta h)^2]$
for $\varepsilon =0$. 

The derivation of Eqs. (\ref{incoh}), (\ref{coh}), (\ref{tauc1}), and (\ref{tauc2}),
from which we digressed below Eq. (\ref{aoft}) is now resumed.
It is assumed that $h(t)$ follows a stationary random Gaussian process, centered on $0$
[a non-zero value simply redefines $\varepsilon$
in Eq. (1)], and $h(t)$ is correlated over time, with a correlation
time $\tau_h$ defined by
$\tau_h\langle  h^2 \rangle=\int_0^{\infty} \langle h(t^\prime) h(t^\prime +t)\rangle_h dt$, 
where $\langle \ldots \rangle_h$ stands for an average over all $h(t)$ histories.
The probability,
\begin{equation}
P(t)=\Delta^2\int_0^t dt_2\int_0^{t_2}dt_1\{ e^{2i\int_{t_1}^{t_2}[h(\tau )-\varepsilon ]d\tau}+cc\},
\label{poft}
\end{equation}
follows straightforwardly from Eq. (\ref{aoft}) and the statement following it.
Now, $\int _{t_1}^{t_2}h(\tau )d\tau$ follows a Gaussian distribution
since $h(t)$ follows, by assumption, from a stationary random Gaussian process.
Therefore, after averaging over all histories of $h(t)$, Eq. (\ref {poft}) becomes,
\begin{equation}
P(t)=\Delta ^2\int_0^t dt_2\int_0^{t}dt_1\;e^{-2F(t_1,t_2)}
\cos {[2\varepsilon (t_2-t_1)]},
\label{p}
\end{equation}
where $F(t_1,t_2)=\int_{t_1}^{t_2}d\tau_1
\int_{t_1}^{t_2}d\tau_2\langle  h(\tau_1) h(\tau_2)\rangle_h$.
Now, 
$F(t_1,t_2)\simeq 2  (\delta h)^2 \tau_h |t_2-t_1|$ if
$\tau_h\ll |t_2-t_1|$, 
and a bit of reflection shows that this relation applies if $\tau\delta h\ll 1$
and $t\gg 1/[\tau_h(\delta h)^2]$ (that is, if $t\gg\tau_\phi$), which leads
to Eq. (\ref{coh}). On the other hand, 
$F(t_1,t_2)\simeq (\delta h)^2 |t_2-t_1|^2$ if $\tau_h\gg |t_2-t_1|$,
and this relation is applicable if $\tau_h\delta h\gg 1$ and $t\gg 1/(\tau_h\delta h)$
(that is, if $t\gg \tau_\phi$), which leads to Eq. (\ref{incoh}).
Similarly, 
\begin{equation}
q=\pm {{i\Delta}\over{\surd P(t)}}\int_{0}^t dt_1e^{2i\varepsilon t_1}e^{-2F(0,t_1)},
\end{equation}
follows from Eq. (\ref{qq}) and the assumption of a stationary random Gaussian $h(t)$.
By the same arguments that led to Eqs. (\ref{incoh}), and (\ref{coh}) above, equations
(\ref{tauc1}) and (\ref{tauc2}) follow straightforwardly, independently of $\varepsilon$.

Numerical results obtained for $\Gamma$ and for $\tau_\phi$ are shown in Fig. 3.
There is good agreement with Eqs. (\ref{incoh}), (\ref{coh}),  (\ref{tauc1}),
and  (\ref{tauc2}). 

In conclusion, coherence effects have been shown to set in
when the coherence time $\tau_\phi$ is longer than the correlation time $\tau_h$
for $h(t)$, and that this comes about because decoherence
effects that arise from $h(t)$ partially average out when
$\tau_h\lesssim 1/\delta h$. Coherence times are then larger than $\tau_h$.
Equations (\ref{incoh}) and (\ref{coh})
which give $\tau_\phi$ and the tunneling rate as a function
of $\tau_h\delta h$ both for $\tau_h\delta h\gtrsim\hbar$,
where the theory of Prokof'ev and Stamp\cite{stamp1} hold, and for $\tau_h\delta h\lesssim\hbar$,
where their theory does not hold, are derived.
 
The condition $\tau_h\delta h\lesssim 1$, which gives $\tau_\phi\gtrsim\tau_h$,
would be fulfilled in spin tunneling systems in which nuclear spin motion
is driven by other atoms in the system with which hyperfine interaction
strengths $\delta^\prime$ are larger than $\delta$ with the spins of interest 
if $\tau_h^{-1}\sim \delta^\prime$. There seems to be some indirect
experimental evidence that
$\tau_\phi\gtrsim\tau_h$, or, equivalently, that $\tau_h\delta h\lesssim 1$
in Fe$_8$ and in Mn$_{12}$ clusters. 
Resolution of the tunneling pair of energy eigenstates has
been reported in Fe$_8$ \cite{tejada} and in Mn$_{12},$\cite{gatto}
under large transverse fields, both at $680$ MHz.
This implies that $\tau_\phi\gtrsim 10^{-9}$ seconds in both cases.
This is to be compared with the value of $1/\delta h$ that $\tau_\phi$ would
equal to [see, Eqs. (\ref{dH}) and (\ref{tauc1})] if $\tau_h\delta h\gtrsim 1$ were fulfilled.
Now, $\delta h\approx 0.02$K for Fe$_8$ (as follows from
measured hyperfine fields of $10^{-3}$T reported by Wernsdorfer\cite{werns1},
$g=2$, and $S=10$), that is, $1/\delta h\approx 0.5\times 10^{-9}$s.
For Mn$_{12}$, $\delta h\approx 10^{-2}$T, (see Ref. \cite{luis2}),
from which $1/\delta h\approx 0.5\times 10^{-10}$s follows. It follows
that, in fact, $\tau_h\delta h\gtrsim 1$ is {\it not} fulfilled, neither
for Fe$_8$ nor for Mn$_{12}$,
suggesting that tunneling in Fe$_8$ and in Mn$_{12}$ proceeds with
some degree of coherence, even when no external transverse field is applied.
\cite{fer}

Enlightening discussions with Dr. F. Luis and with Professor J. Villain
have been of much help.
A research grant, No. PB98-0541, from DGESIC of Spain is gratefully acknowledged.

\end{multicols}

\begin{references}

\bibitem{nature1}See, for instance, 
B. Schwarzschild, in {\it Physics Today}, {\bf 50}, (1), 17 (1997).

\bibitem{zurek}W. H. Zurek, Phys. Rev. D {\bf 24}, 1516 (1981);
W. H. Zurek, Physics Today {\bf 44}, No. 10, 36 (1991);
J. P. Paz and W. H. Zurek, Phys. Rev. Lett. {\bf 82}, 5181 (1999);
D. Giulini, E. Joos, C. Kiefer, J. Kupusch, I. -O. Stamtescu, and H. D. Zeh,
{\it Decoherence and the appearance of a classical world in quantum theory}
(Springer, Heidelberg, 1996).

\bibitem{unruh}W. G. Unruh, Phys. Rev. A {\bf 51}, 992 (1995); for more recent work
and further references, see, for instance,
L. Viola and S. Lloyd, Phys. Rev. Lett. {\bf 82}, 2417 (1999). 

\bibitem{luis2}F. Luis, J. Bartolom\'e, and J. F. Fern\'andez,
Phys. Rev. B {\bf 57}, 505 (1998).

\bibitem{julio}J. F. Fern\'andez, F. Luis, and J. Bartolom\'e,
Phys. Rev. Lett. {\bf 80}, 5659 (1998); for a different viewpoint
see, J. Villain, A. W\"urger, A. Fort, A. Rettori, J. Phys.
(Paris) {\bf 7}, 1583 (1997); A. Fort, A. Rettori, J. Villain, D. Gatteschi, and
R. Sessoli, Phys. Rev. Lett. {\bf 80}, 612 (1998).

\bibitem{wern0}For a brief review, see W. Wernsdorfer and R. Sessoli,
Science {\bf 284}, 133 (2000).

\bibitem{garg}A. Garg, Phys. Rev. Lett. {\bf 74}, 1458 (1995).

\bibitem{stamp1}N. V. Prokof'ef and P. C. E. Stamp,J. Low Temp. Phys.,
{\bf 104}, 143 (1996).

\bibitem{stamp2}N. V. Prokof'ef and P. C. E. Stamp, Phys. Rev. Lett. {\bf 80},
5794 (1998). 

\bibitem{nikitin}L. Landau, Phys. Z. Sowjetunion, {\bf 2}, 46 (1932);
C. Zener, Proc. R. Soc. London, A{\bf 137}, 696 (1932);
see also, for instance, E. E. Nikitin and L. Z\"ulicke, in
{\it Theory of Chemical Elementary Processes},
edited by, G. Berthier, et al. (Springer-Verlag, New York, 1978);
B. H. Bransden, {\it Atomic Collision Theory} (The Benjamin Cummings Publishing
Company, Inc., Reading, MA, 1983).

\bibitem{leggett}A. J. Leggett, S. Chakravarty, A. T. Dorsey, M. P. A. Fisher,
A. Garg, and, W. Zwerger, Rev. Mod. Phys. {\bf 59}, 1 (1987);
K. M. Forsythe and N. Makri, Phys. Rev. B {\bf 60}, 972 (1999).

\bibitem{referee1}For a different measure of coherence, see, for instance,
A. Tameshtit and J. E. Sipe, Phys. Rev. A {\bf 47}, 1697 (1993).

\bibitem{tejada}E. Del Barco, N. Vernier, J. M. Hern\'andez, J. Tejada, E. M.
Chudnovsky, E. Molins, and G. Bellessa, Europhys. Lett. {\bf 47}, 722 (1999).

\bibitem{gatto}G. Bellessa, N. Vernier, B. Barbara, and D. Gatteschi,
Phys. Rev. Lett. {\bf 83}, 416 (1999).

\bibitem{werns1}W. Wernsdorfer, T. Ohm, C. Sangregorio, R. Sessoli,
D. Mailly, and C. Paulsen, Phys. Rev. Lett. {\bf 82}, 3903 (1999).

\bibitem{fer}A relaxation time $T_1>10^{-7}$s is obtained in 
ac specific heat experiments, F. Luis, F. Mettes, J. Tejada, D. Gatteschi,
and L. J. de Jongh Phys. Rev. Lett. xxx, xx (2000), but $\tau_\phi <T_1$ is expected.

\end{references}
\end{document}